\documentclass[12pt,onecolumn]{IEEEtran}

\IEEEoverridecommandlockouts
\usepackage{cite}
\usepackage{amsmath,amssymb,amsfonts,accents,mathrsfs}
\usepackage{textcomp}
\usepackage{nicefrac}
\def\BibTeX{{\rm B\kern-.05em{\sc i\kern-.025em b}\kern-.08em
    T\kern-.1667em\lower.7ex\hbox{E}\kern-.125emX}}   

\usepackage{float}
\usepackage{bm}

\usepackage{graphicx}
\usepackage{textcomp}
\usepackage[dvipsnames]{xcolor}

\definecolor{amethyst}{rgb}{0.6, 0.4, 0.8} 

\usepackage{tikz}
\usetikzlibrary{positioning}
\usepackage{textcomp}
\usetikzlibrary{shapes,arrows,backgrounds}
\usetikzlibrary{datavisualization}
\usetikzlibrary{patterns}
\usepackage{verbatim}

\usepackage{pgfplots} 
\pgfplotsset{compat=newest} 
\pgfplotsset{plot coordinates/math parser=false} 
\newlength\figureheight 
\newlength\figurewidth 

\definecolor{x11_gray}{rgb}{0.85, 0.85, 0.85}
\definecolor{darkgreen}{rgb}{0.0, 0.5, 0.0}
\definecolor{amethyst}{rgb}{0.6, 0.4, 0.8}

\usepackage[labelsep=quad,indention=10pt]{subfig}
\usepackage{algorithm}
\usepackage{algorithmic}

\usepackage{array}

\usepackage{amsmath,amssymb,amsfonts}
\usepackage{epstopdf,epsfig}
\usepackage{cite}

\renewcommand{\i}{ {\left[i\right]}}
\renewcommand{\ni}{ {\left[i\right]}}

\begin{document}

\title{{\LARGE{Study of Adaptive Activity-Aware Iterative Detection Techniques for Massive Machine-Type Communications}}}

\author{Roberto~B.~Di~Renna,~\IEEEmembership{Student Member,~IEEE,}
        and~\\ Rodrigo~C.~de~Lamare,~\IEEEmembership{Senior Member,~IEEE}
\thanks{The authors are with the Centre for Telecommunications Studies (CETUC),
Pontifical Catholic University of Rio de Janeiro (PUC-Rio), Rio de Janeiro 22453-900,
Brazil (e-mail: robertobrauer@cetuc.puc-rio.br; delamare@cetuc.puc-rio.br).
This work was supported by Conselho de Nacional de Desenvolvimento Cientifico e Tecnologico (CNPq) grant, funded by the Brazilian government.}
\thanks{}}


\maketitle

\begin{abstract}
This work studies the uplink of grant-free low data-rate massive
machine-to-machine communications (mMTC) where devices are only
active sporadically, which requires a joint activity and data
detection at the receiver. We develop an adaptive decision feedback
detector along with an $l_0$-norm regularized activity-aware
recursive least-squares algorithm that only require pilot symbols.
An iterative detection and decoding scheme based on low-density
parity-check (LDPC) is also devised for signal detection in mMTC.
Simulations show the performance of the proposed approaches against
existing schemes.
\end{abstract}


%
\IEEEpeerreviewmaketitle

\section{Introduction} \label{sec:intro}

\IEEEPARstart{T}{he} fifth generation (5G) of mobile radio
communication systems should support Internet of Things (IoT),
Internet of Everything (IoE), and Industry 4.0, which are referred
to as massive machine type communications (mMTC). In contrast to
human-type communications (HTC) the traffic of mMTC originates from
a large number of devices concentrated in the uplink with small
packets of up to a few hundred bits that transmit
sporadically~\cite{Chen2017}. Due to the huge number of mMTC devices
it is impractical to pre-allocate resources to them. Therefore, the
solution is to employ grant-free random access. As the number of
active mMTC devices is random their average value should be
considered and the network should maximize the arrival rate that can
be supported given the available radio resources. Furthermore, the
packet error rate of a unique mMTC transmission is on the order of
$10^{-1}$~\cite{Popovski2018} and each packet is structured in a
preamble (metadata) and a payload (data)~\cite{Liu2018}. In
Low-Active Code Division Multiple Access (LA-CDMA) schemes, metadata
are used as the ID of each user and, due to the huge number of
devices, it is infeasible to provide orthogonal metadata (pilot
symbols) for channel estimation and receive processing of active
users ~\cite{Azari2017}.

Prior work on channel estimation and detection of devices includes
compressed sensing (CS) methods that outperform conventional channel
estimation techniques~\cite{JChoi2017}. However, if the device
activity detection is accurate then conventional channel estimation
techniques may be employed. As device activity detection is an open
problem, most of the works to date consider perfect channel state
information (CSI) at the BS and investigate different detection
approaches~\cite{Knoop2014,Ahn2018,DiRenna2019}. Unlike prior work
with successive interference cancellation and list-based detectors
\cite{deLamare2003},\cite{itic},\cite{deLamare2008},\cite{cai2009},\cite{jiomimo},\cite{dfcc},\cite{deLamare2013},\cite{did},\cite{rrmser}
detectors for mMTC must exploit the activity of devices.

In this work, we propose a decision feedback detector along with an
activity-aware recursive least squares (RLS) algorithm (AA-RLS-DF)
based on an $l_0$-norm penalty function that does not require
explicit channel estimation. The regularization exploits the
sparsity in the coefficients update formulation, thus improving the
performance of parameter estimation and detection. An iterative
detection and decoding (IDD) scheme based on low-density
parity-check (LDPC) is also devised for signal detection. Unlike
existing approaches, the proposed AA-RLS-DF scheme exploits the
activity of devices. Simulations show that the proposed AA-RLS-DF
scheme significantly outperforms prior work with a competitive
complexity.

The organization of this paper is as follows:
Section~\ref{sec:syst_model} briefly describes the LA-CDMA system
model and the augmented alphabet specifically designated for mMTC.
Section~\ref{sec:ite_det_dec} describes the adaptive implementation,
the $l_0$ regularization and introduces the iterative detection and
decoding, proposed for mMTC. Section~\ref{sec:sim} presents the
setup for simulations and the results while Section~\ref{sec:Conc}
draws the conclusions.

\textit{Notation:} Matrices and vectors are denoted by boldfaced
capital letters  and lower-case letters, respectively. The space of
complex (real) $N$-dimensional vectors is denoted by
$\mathbb{C}^N\left(\mathbb{R}^N\right)$. The $i$-th column of a
matrix $\mathbf{A} \in \mathbb{C}^{M\times N}$ is denoted by
$\mathbf{a}_i \in \mathbb{C}^M$. The superscripts
$\left(\cdot\right)^\text{T}$ and $\left(\cdot\right)^\text{H}$
stand for the transpose and conjugate transpose, respectively. For a
given vector $\mathbf{x} \in \mathbb{C}^N, ||\mathbf{x}||$ denotes
its Euclidean norm. $\mathbb{E\left[\cdot\right]}$ stands for
expected value and $\mathbf{I}$ is the identity matrix.

\section{System model} \label{sec:syst_model}

We assume the uplink of slot-synchronized mMTC devices in which the
base station (BS) with multiple antennas receives data from multiple
devices with a single antenna. The transmitted data, at each time
instant, is composed by symbols from a regular modulation scheme
denoted by $\mathcal{A}$, as quadrature phase-shift keying (QPSK) if
the device is active, or zero, otherwise. So, each transmitted
symbol belongs to the augmented alphabet $\mathcal{A}_0$, which
consists of a regular modulation scheme and the zero ($\mathcal{A}_0
= \mathcal{A}\cup \left\{0\right\}$).

\begin{figure*}[t]
    \begin{center}
        \includegraphics[scale=0.4]{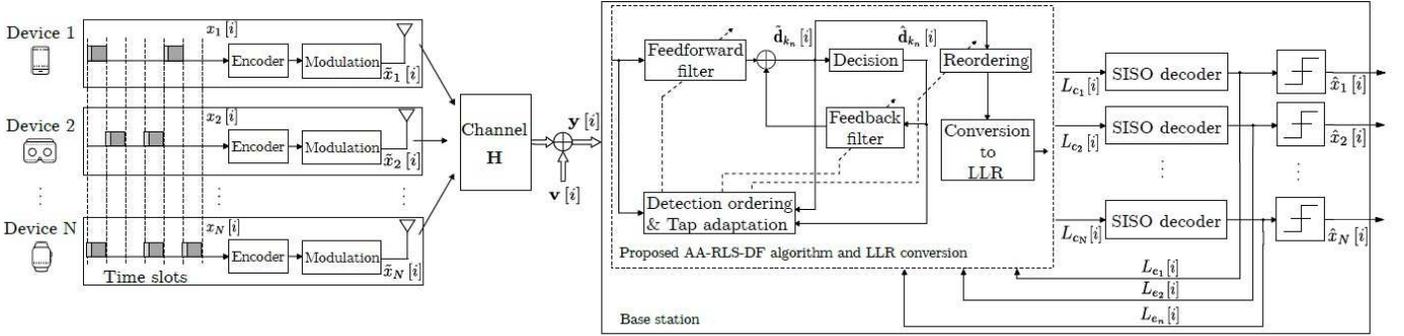}
        \caption{Block diagram of AA-RLS-DF with IDD structure.}
        \label{fig:dfe_idd_structure}
    \end{center}
\end{figure*}

In the mMTC system of Fig.~\ref{fig:dfe_idd_structure}, the data
symbols of the $N$ devices are transmitted over Rayleigh fading
channels, grouped into a $M \times N$ channel matrix $\mathbf{H}$
which contains the spreading sequences and channel impulse responses
to the BS. The received signal is collected into an $M \times 1$
vector $\mathbf{y}\ni$ as
    \begin{equation}
       \mathbf{y} {\left[i\right]} = \mathbf{H} \mathbf{x} {\left[i\right]} + \mathbf{v} {\left[i\right]}
    \end{equation}
\noindent where $i$ is the time index, the $M \times 1$ vector
$\mathbf{v}\ni$ is a zero mean complex circular symmetric Gaussian
noise with covariance matrix
$\mathbb{E}\left[\mathbf{v}\ni\mathbf{v}\ni^\text{H}\right] =
\sigma^2_v \mathbf{I}$. The $N \times 1$ symbol vector ${\bf x}\ni =
[x_1\ni \ldots x_k\ni \ldots x_N\ni]^\text{T}$ has zero mean and
covariance matrix
$\mathbb{E}\left[\mathbf{x}\ni\mathbf{x}\ni^\text{H}\right] =
\sigma^2_x \mathbf{I}$, where $\sigma_x^2$ is the signal power of
the active devices.

Since in the mMTC scenario devices have low-activity probability the
system can be interpreted as a sparse signal processing system
model. In this way, even in an underdetermined system ($N >> M$), it
is possible to recover the transmitted vector $\mathbf{x}\ni$. In
order to avoid the need to estimate channels, we propose an adaptive
scheme which considers the activity of devices, combined with an
$l_0$-norm penalty function.

\section{Proposed Iterative Detection and Decoding} \label{sec:ite_det_dec}

The proposed AA-RLS-DF detector is introduced in this section along
with the detection ordering algorithm based on the least squares
criterion (LSE). We also detail the proposed $l_0$-norm regularized
RLS parameter estimation algorithm that exploits sparsity to refine
the detection and estimation tasks. Lastly, we present an IDD scheme
based on the proposed AA-RLS-DF detector and LDPC codes.

\subsection{Adaptive Detection Design} \label{subsec:adapt_imp}

{ {In order to reduce detection errors, we propose an adaptive
AA-RLS-DF detector, which performs parameter estimation using
metadata in training mode and then carries out data detection in a
decision-directed mode.  The receive filters and the ordering are
updated at each iteration, as illustrated in Fig. 1. The feedforward
and feedback receive filters are written as}}

\vspace{-2.5pt}
    \begin{equation}\label{eq:filter_forw_fedd}
        \mathbf{w}_{n}\left[i\right] =
        \left\{
        \begin{array}{cl}
            \mathbf{w}^{f}_{n}\left[i\right], & n=1; \\
            \left[{\mathbf{w}^{f}_{n}}^\text{T}\left[i\right],{\mathbf{w}^{b}_{n}}^\text{T}\left[i\right]\right]^\text{T}, &    n=2,\dots, N.
        \end{array}\right.
    \end{equation}

and the augmented received vector,
\vspace{-2.5pt}
    \begin{equation}\label{eq:all_y}
        \mathbf{y}_{n}\left[i\right] =
        \left\{
        \begin{array}{cl}
            \mathbf{y}\left[i\right], & n=1; \\
            \left[{\mathbf{y}}^\text{T}\left[i\right],{\hat{\mathbf{d}}^\text{T}_{\phi_{n-1}}}\left[i\right]\right]^\text{T}, & n=2,\dots, N.
        \end{array}\right.
    \end{equation}

\noindent  where $\mathbf{w}_{n}\left[i\right] =
\left[w^{f}_{n,1}\left[i\right], \dots, w^{f}_{n,M}\left[i\right],
w^{b}_{n,M+1}\left[i\right],\dots,\right.$\\ $\left.
w^{b}_{n,M+N}\left[i\right]\right]^\text{T}$ { {corresponds to both
filters used for the detection of the symbol of the $n$-th device
and the $N \times 1$ vector
$\hat{\mathbf{d}}_{\phi_{n}}\left[i\right]$ contains the previously
detected symbols of the $n$-th device. The soft symbol estimates of
the $n$-th device are obtained by}}

    \begin{equation}\label{eq:out_equa}
        \tilde{d}_{\phi_n}\left[i\right] = \mathbf{w}_n^\text{H}\left[i\right]\mathbf{y}_n\left[i\right].
    \end{equation}

To determine the detection order $\phi_n$, the set of candidate
symbols are evaluated. Then, we select the symbols associated with
the minimum LSE. The receive filter is computed by the $l_0$-norm
regularized RLS, whose cost function is given by \vspace{-5.5pt}
    \begin{eqnarray}\label{eq:cost_func_rls}
    \mathcal{J}_{n,j}\left[i\right] &=&  \sum_{l=0}^{i} \lambda^{i-l} \left|\hat{d}_{j}\left[l\right]-\mathbf{w}_n^\text{H}\left[i\right]\mathbf{y}_n\left[i\right]\right|^2,
    \end{eqnarray}

\noindent where $0< \lambda \leq 1$ is the forgetting factor which
gives exponentially less weight to older error samples. { {At each
symbol detection, the index of the chosen feedback filter is stored
in  the sequence of detection $\phi_n$, represented as

\vspace{-2.5pt}
    \begin{equation}\label{eq:detec_ord}
        \phi_n = \underset{j\in S_i}{\textrm{arg min}}\hspace*{5pt} \mathcal{J}_{n,j}\left[i\right].
    \end{equation}

After this, the filter weights are updated and the result of the decision block, given by

    \begin{equation}\label{eq:out_equa2}
        \hat{d}_{\phi_n}\left[i\right] = \mathcal{Q}\left[\mathbf{w}_n^\text{H}\left[ {i}\right]\mathbf{y}_n\left[i\right]\right].
    \end{equation}

\noindent The received vector is concatenated with the output
of~(\ref{eq:out_equa2}), as in~(\ref{eq:all_y}). In the next
detection, the new cost functions are computed just for the symbols
which have not been detected yet. For this, index $j$
in~(\ref{eq:cost_func_rls}) belongs to the set
$S_i=\{1,2,\dots,N\}-\{\phi_1,\phi_2,\dots,\phi_{n-1}\}$, which
contains the index of not yet detected symbols.}}

{ {After using the metadata to adjust the receive filters, AA-RLS-DF
starts the decision-directed mode. The received data $\mathbf{y}\i$,
initially repeats the same sequence of operations in the adaptation
mode but, after obtaining all soft estimates $\tilde{\mathbf{d}}\i$,
the vector is reorganized in the original sequence and each soft
estimate is converted to a log-likelihood ratio (LLR) ($L_{c_n}\i$).
After the SISO decoder block, the extrinsic LLRs ($L_{e_n}\i$)
return in the flow to the next IDD iteration, refining the new
$L_{c_n}\i$ computation. }}


\subsection{$l_0$-norm Regularized RLS Algorithm} \label{subsec:reg_alg}

In order to exploit the sparse activity of devices and compute the
parameters of the proposed DF detector without the need for explicit
channel estimation, we devise an $l_0$-norm regularized RLS
algorithm that minimizes the cost function: \vspace{-2.5pt}
    \begin{equation} \label{eq:cost_func_rls_l0}
    \mathcal{J}_{n,j}\left[i\right] = \sum_{l=0}^{i} \lambda^{i-l} \left|\hat{d}_{j}\left[l\right]-\mathbf{w}_n^\text{H}\left[i\right]\mathbf{y}_n\left[i\right]\right|^2 + \, \gamma \|\mathbf{w}_{n}\left[i\right]\|_0
    \end{equation}

\noindent where $\|\cdot\|_0$ denotes $l_0$-norm that counts the
number of zero entries in $\mathbf{w}_{n}$ and $\gamma$ is a
non-zero positive constant to balance the regularization and,
consequently, the estimation error. We consider an approximation to
the regularization term
\cite{jidf},\cite{Angelosante2010},\cite{l1stap},\cite{saalt}.

Approximating the value of the $l_0$-norm~\cite{Bradley98}, the cost
function in (\ref{eq:cost_func_rls_l0}) can be rewritten as
    \vspace{-2.5pt}
    \begin{eqnarray}\label{eq:cost_func_rls_l0_app}
    \mathcal{J}_{n,j}\left[i\right] &=& \sum_{l=0}^{i} \lambda^{i-l} \left|\hat{d}_{j}\left[l\right]-\mathbf{w}_{n}^\text{H}\left[i\right]\mathbf{y}_n\left[i\right]\right|^2 + \gamma \sum^{2M}_{p=1} \left(1-\text{exp}\left(-\beta|w_{n,p}\left[i\right]|\right)\right).
    \end{eqnarray}

The parameter $\beta$ regulates the range of the attraction to zero
on small coefficients of the filter. Thus, taking the partial
derivatives for all entries $i$ of the coefficient vector
$\mathbf{w}_{n}\i$ in (\ref{eq:cost_func_rls_l0_app}) and setting
the results to zero, yields
    \vspace{-0.5pt}
    \begin{eqnarray}\label{eq:recursive_rls_l0}
        \mathbf{w}_{n}\left[i\right] &=& \mathbf{w}_{n}\left[i-1\right] + \mathbf{k}\left[i\right] \epsilon_n^\ast\left[i\right] - \gamma \, \beta\, \text{sgn}\left(w_{n,p}\left[i\right]\right)\text{exp}\left(-\beta|w_{n,p}\left[i\right]|\right)
    \end{eqnarray}

\noindent where { {$\mathbf{k}\left[i\right]$ is the gain vector}}
and $\text{sgn}\left(\cdot\right)$ is a component-wise sign function
defined as
    \vspace{-0.5pt}
    \begin{equation}
        \text{sgn}\left(w_{n,p}\left[i\right]\right) =
        \left\{
        \begin{array}{rl}
            w_{n,p}\left[i\right]/|w_{n,p}\left[i\right]|, & w_{n,p}\left[i\right] \neq 0; \\
            0,                           & \text{otherwise.}
        \end{array}\right.
    \end{equation}

    In order to reduce computational complexity in (\ref{eq:recursive_rls_l0}), the exponential function is approximated by the first order of Taylor series expansion, given by
\vspace{-0.5pt}
    \begin{equation}\label{eq:exp_app}
        \text{exp}\left(-\beta\, |w_{n,p}\left[i\right]|\right) \approx
        \left\{
        \begin{array}{rl}
            1-\beta\, |w_{n,j}\left[i\right]|, & |w_{n,j}\left[i\right]| \leq 1/\beta; \\
            0,                           & \text{otherwise.}
        \end{array}\right.
    \end{equation}

    As the exponential function is positive, the approximation of (\ref{eq:exp_app}) is also positive. In this way, (\ref{eq:recursive_rls_l0}) becomes
\vspace{-2.5pt}
    \begin{eqnarray}\label{eq:recursive_rls_l0_app}
        \mathbf{w}_{n}\left[i\right] &=& \mathbf{w}_{n}\left[i-1\right] + \mathbf{k}\left[i\right] \epsilon_n^\ast\left[i\right]  - \gamma \, \beta\, \text{sgn}\left(w_{n,p}\left[i\right]\right)f_\beta\left(w_{n,p}\left[i\right]\right)
    \end{eqnarray}

\noindent where the function $f_\beta\left(w_{n,p}\left[i\right]\right)$ is given by

    \begin{equation}\label{eq:f_beta}
        f_\beta\left(w_{n,p}\left[i\right]\right) =
        \left\{
        \begin{array}{rl}
            \beta^2 \left(w_{n,p}\left[i\right]\right)+\beta, & -1/\beta \leq w
            _{n,p}\left[i\right]< 0; \\
            \beta^2 \left(w_{n,p}\left[i\right]\right)-\beta, & 0< \leq w_{n,p}\left[i\right]\leq 1/\beta; \\
            0,                           & \text{otherwise.}
        \end{array}\right.
    \end{equation}

    We notice that the function $f_\beta\left(w_{n,j}\left[i\right]\right)$ in (\ref{eq:recursive_rls_l0_app}) imposes an attraction to zero in small coefficients. So, if the value of $w_{n,j}\left[i\right]$ is not equal or between to $\left[-1/\beta,1/\beta\right]$, no additional attraction is exerted. Thus, the convergence rate of near-zero coefficients of parameters of devices in mMTC applications that exhibit sparsity will be increased~\cite{Bradley98}. { {The computational cost of AA-RLS-DF is $O(N^2)$, which is comparable with a standard DF detector with an RLS algorithm. Since the other considered algorithms have a computational cost of $O(N^3)$, AA-RLS-DF outperforms both performance and efficiency.}} The pseudo-code, which also considers an IDD scheme with AA-RLS-DF, is described in Algorithm 1.

\subsection{Proposed Soft Information Processing and Decoding}\label{subsec:prop_it_det_dec}

In this section, the structure of the proposed IDD scheme
\cite{aaidd} is described. Unlike prior IDD schemes
\cite{XWang1999}, \cite{bfidd}, \cite{1bitidd} which employ
convolutional, Turbo and LDPC \cite{dopeg,memd} codes, our scheme
does not require matrix inversions and exploit the activity of
devices to refine the iterative processing. We estimate and
incorporate the probability of each device being active in the mMTC
scenario in the computation of each a priori probability symbol.

Based on $L_{e_n}$, provided by the LDPC decoder and assuming the
bits are statistically independent of one another \cite{vfap}, the
\textit{a priori} probabilities are calculated as \vspace{-0.5pt}
    \begin{equation}\label{eq:a_priori1}
        \text{Pr}\left(x_n\i = x\right) = \sum_{x\in \mathcal{A}_0} x\left(\prod^{M_c}_{z=1}\left[1+\text{exp}\left(-x^z L_{e_n}^z\i\right)\right]^{-1}\right),
    \end{equation}

{ {\noindent where $M_c$ represents the total number of bits of
symbol $x$, the superscript $z$ indicates the $z$-th bit of symbol
of $x$, in $x^z$ (whose value is $(+1,-1)$) and the form of
extrinsic LLR, $L_{e_n}^z\i$. As $\mathcal{A}_0$ is the augmented
alphabet, the considered symbols are from the modulation scheme
chosen and zero. Considering the probability of the $n$-th device
being active as $\rho_n$, the \textit{a priori} probabilities can be
rewritten as}}

\vspace{-0.5pt}
\begin{align}
    \text{Pr}\left(x_n\i = x\right) \left\{
    \begin{array}{rl}
     \hspace{-4pt}\rho_n + \left(1-\rho_n\right)\text{Pr}\left(x_n\i = x\right), & \hspace{-7pt} {\scriptstyle{\text{if}\, \left(x^1 \text{ and } x^2\right) = 0,}}   \\ \label{eq:a_priori2}
      \left(1-\rho_n\right)\text{Pr}\left(x_n\i = x\right), & \hspace{-7pt} \text{{\small{otherwise}}}.
    \end{array}\right.
\end{align}


Eq.(\ref{eq:a_priori2}) incorporates the \textit{a priori}
probabilities related to the probability that the $n$-th device is
active. { {We consider different $\rho_n$ values for each user,
drawn uniformly at random in the interval $[0.1, 0.3]$}}. Due to the
large number of independent variables considered to compute the
filter output, it can be approximated by a complex Gaussian
distribution~\cite{XWang1999}. Hence, we approximate
$\hat{d}_{\phi_n}\i$ by the output of an equivalent AWGN channel
with $\hat{d}_{\phi_n}\left[i\right] = \mu_n\left[i\right]
x_n\left[i\right] + b_n\left[i\right]$, where \vspace{-0.5pt}
    \begin{eqnarray}\label{eq:mean_gauss_app} \nonumber
        \mu_n\left[i\right] &=& \mathbb{E}\left\{\hat{d}_{\phi_n}\left[i\right]x_{n}\left[i\right]\right\}= \mathbb{E}\left\{\mathbf{w}_n^\text{H}\left[i\right]\mathbf{y}_n\left[i\right]x_{n}\left[i\right]\right\} \\
        & {\approx}& \mathbf{w}_n^\text{H}\left[i\right] \left(\sum_{p=1}^{i}\lambda^{i-p}\, \mathbf{y}_n\left[p\right]x_{n}\left[p\right]\right). \\ \nonumber
    \end{eqnarray}
\vspace{-1.5pt}
Note that $x_n$ are the metadata symbols and $b_n\left[i\right]$ are zero-mean complex Gaussian variables with variance $\eta^2_n\left[i\right]$ as
\vspace{-0.5pt}
    \begin{eqnarray}\label{eq:var_gauss_app} \nonumber
        \eta_n^2\left[i\right] &=& \text{var}\left\{\hat{d}_{\phi_n}\left[i\right]\right\} = \mathbb{E}\left\{\|\hat{d}_{\phi_n}\left[i\right]\|^2\right\}-\mu_n^2\left[i\right]  \\
        &=& \mathbf{w}_n^\text{H}\left[i\right]\mathbb{E}\left\{{\mathbf{y}_n}\left[i\right]\mathbf{y}_n^\text{H}\left[i\right]\right\}\mathbf{w}_n\left[i\right] - \mu_n^2\left[i\right] \\ \nonumber
        & {\approx}& \mathbf{w}_n^\text{H}\left[i\right]\left(\sum_{p=1}^{i}\lambda^{i-p}\, \mathbf{y}_n\left[p\right]\mathbf{y}_n^\text{H}\left[p\right]\right)\mathbf{w}_n\left[i\right] - \mu_n^2\left[i\right].
    \end{eqnarray}

Then, the extrinsic LLR computed by the AA-RLS-DF detector for the $z$-th bit ($z \in \{1,\dots,M_c\}$) of the symbol $x_n$ transmitted by the $n$-th device is

\begin{align}\label{eq:llr}
        L_{c_n}^z\i =& \log \frac{\sum_{x\in \mathcal{A}_z^{+1}}\text{Pr}\left(\hat{d}_{\phi_n}\i|\, x\right)\text{Pr}\left(x\right)}{\sum_{x\in \mathcal{A}_z^{-1}}\text{Pr}\left(\hat{d}_{\phi_n}\i|\, x\right)\text{Pr}\left(x\right)} - L_{e_n}^z\i
\raisetag{20pt}
\end{align}

\noindent where $\mathcal{A}_z^{+1}$ is the set of $2^{M_c -1}$ hypotheses of $x$ for which the $z$-th bit is +1 (analogously for $\mathcal{A}_z^{-1}$). Therefore, the likelihood function $P\left(\hat{d}_{\phi_n}|x\right)$ is approximated by

    \begin{equation}\label{eq:likelihood}
        P\left(\hat{d}_{\phi_n}\i|\, x\right) \approx \frac{1}{\pi\, \eta^2_n\i} \text{exp}\left(-\frac{1}{\eta_n^2\i}|\hat{d}_{\phi_n}\i-\mu_n\i x|^2\right).
    \end{equation}

\begin{table}[h]
    \begin{center}
        \begin{tabular}{ll} \\ \hline
        \multicolumn{2}{l}{\textbf{Algorithm 1} Proposed IDD with AA-RLS-DF algorithm} \\ \hline
        1.  & Initialization: $M$, $N$, $\bm{\rho}$, $\beta$, $\gamma$, $\lambda$, $\mathbf{P}_n = \bm{\rho}\,\mathbf{I}_{\text{M}}$\\
        2.  & Compute the \textit{a priori probability} with Eqs.~(\ref{eq:a_priori1}) and (\ref{eq:a_priori2});  \\
            & \footnotesize \% For each metadata sequence $\hat{\mathbf{d}}_p\i$ and received vector $\mathbf{y}_{p,n}\i$, \\
        \small
        3.  & \hspace{10pt} Compute the Kalman gain vector\\
            & \hspace{10pt} $\mathbf{k}_n\i = (\mathbf{P}_n\i \,\mathbf{y}_{p,n}\i)/(\lambda+\mathbf{y}_{p,n}^\text{H}\i\, \mathbf{P}_n\i \, \mathbf{y}_{p,n}\i)$;\\
        4. & \hspace{10pt} Estimate $\tilde{d}_{\phi_n}\i = \mathbf{w}_n^\text{H}\i\,\mathbf{y}_{p,n}\i$;\\
        5. & \hspace{10pt} Update the error value with $\epsilon_{\phi_n}\i = \hat{d}_{p,\phi_n}\i - \tilde{d}_{\phi_n}\i$;\\
        6. & \hspace{10pt} Update the filters with Eq.~(\ref{eq:recursive_rls_l0_app});\\
        7. & \hspace{10pt} Update the auxiliary matrix\\
           & \hspace{10pt} $\mathbf{P}_n\i = \lambda^{-1} \left(\mathbf{P}_n\i - \mathbf{k}_n\i\, \mathbf{y}_{p,n}^\text{H}\i\,\mathbf{P}_n\i\right)$; \\
        8. & \hspace{10pt} Update the sequence of detection with Eq.(\ref{eq:detec_ord}); \\
           & \footnotesize \% After computing the filter values $\mathbf{w}_{n}\i$, the algorithm starts the\\
           &  \footnotesize direction mode, returning to \textbf{3.} to soft estimate data and go to \textbf{9.}\\
        \small
        9.  & Compute $\mu_{\phi_n}\i$ and $\eta^2_{\phi_n}\i$ with Eqs.~(\ref{eq:mean_gauss_app}) and (\ref{eq:var_gauss_app});\\
        10.  & Verify the likelihood function $P\left(\hat{d}_{\phi_n}\i|x\right)$ with Eq.~(\ref{eq:likelihood}); \\
        11.  & Compute the LLR value according to Eq.(\ref{eq:llr}); \\ \hline
        \end{tabular}
    \end{center}
\end{table}

\section{Simulation Results}\label{sec:sim}

We evaluate the simulation results of an under-determinated mMTC
system with $N=128$ devices and unit-norm random sequences with
length of $M=64$ for spreading. The proposed and existing schemes
experience an independent and identically-distributed (i.i.d.)
random flat-fading channel model and the values are taken from
complex Gaussian distribution of
$\mathcal{C}\mathcal{N}\left(0,1\right)$. The active devices radiate
QPSK symbols with the same power and the activity probabilities
$\left\{\rho_n\right\}^N_{n=1}$ are drawn uniformly at random in
$\left[0.1,0.3\right]$. Each symbol block has 128 symbols, split in
to 60 metadata and 68 data. This balance between pilots and data is
suggested in~\cite{LLiu2018}. All these assumptions are considered
for two scenarios, uncoded and coded systems, in which numerical
results are averaged over $10^5$ runs. The performance of AA-RLS-DF
is compared with other relevant schemes, as the linear mean squared
error (LMMSE), unsorted SA-SIC~\cite{Knoop2014}, SA-SIC with
A-SQRD~\cite{Ahn2018}, AA-MF-SIC~\cite{DiRenna2019} and a version of
AA-RLS-DF without decision feedback, AA-RLS-Linear. As a lower
bound, the Oracle LMMSE detector, which has the knowledge of the
index of nonzero entries, is considered. For all algorithms that
require explicit channel estimation, we considered $\hat{\mathbf{H}}
= \mathbf{H} + \mathbf{E}$, where $\mathbf{H}$ represents the
channel estimate and $\mathbf{E}$ is a random matrix corresponding
to the error for each link. Each coefficient of the error matrix
follows a Gaussian distribution, i.e., $\sim
\mathcal{C}\mathcal{N}\left(0,{\sigma}^2_e\right)$, where
$\sigma^2_e = \sigma_v^2/5$. { {For uncoded systems, the average SNR
is given by $10\log\left(N\, \sigma^2_x/\sigma^2_v\right)$, while
for coded systems is $10\log\left(N R\,
\sigma^2_x/\sigma^2_v\right)$}.

Fig.~\ref{fig:result_wout_code} shows the net symbol error rate
(NSER) which considers only the active devices. LMMSE exhibits poor
performance since the system is under-determined. Due to error
propagation, the unsorted SA-SIC does not perform well. A-SQRD and
AA-MF-SIC are effective since both consider the activity
probabilities, but under imperfect CSI conditions, their performance
is not so good. In contrast, as AA-RLS-DF does not need a explicit
channel estimation,  {it is} more efficient. The decision-feedback
scheme provides a NSER gain due to the interference cancellation.
For the coded systems with IDD, Fig.~\ref{fig:ldpc} shows the BER of
the already considered algorithms and scheme of Wang and
Poor~\cite{XWang1999}. The sparsity of mMTC approach degrades the
expected efficiency of LMMSE-PIC, obtaining little variation in
relation to LMMSE and SA-SIC. The hierarchy of performance of the
other considered algorithms is the same as the uncoded case but with
better error rate values.

\begin{figure}[t]
    \centering
    \includegraphics[scale=0.87]{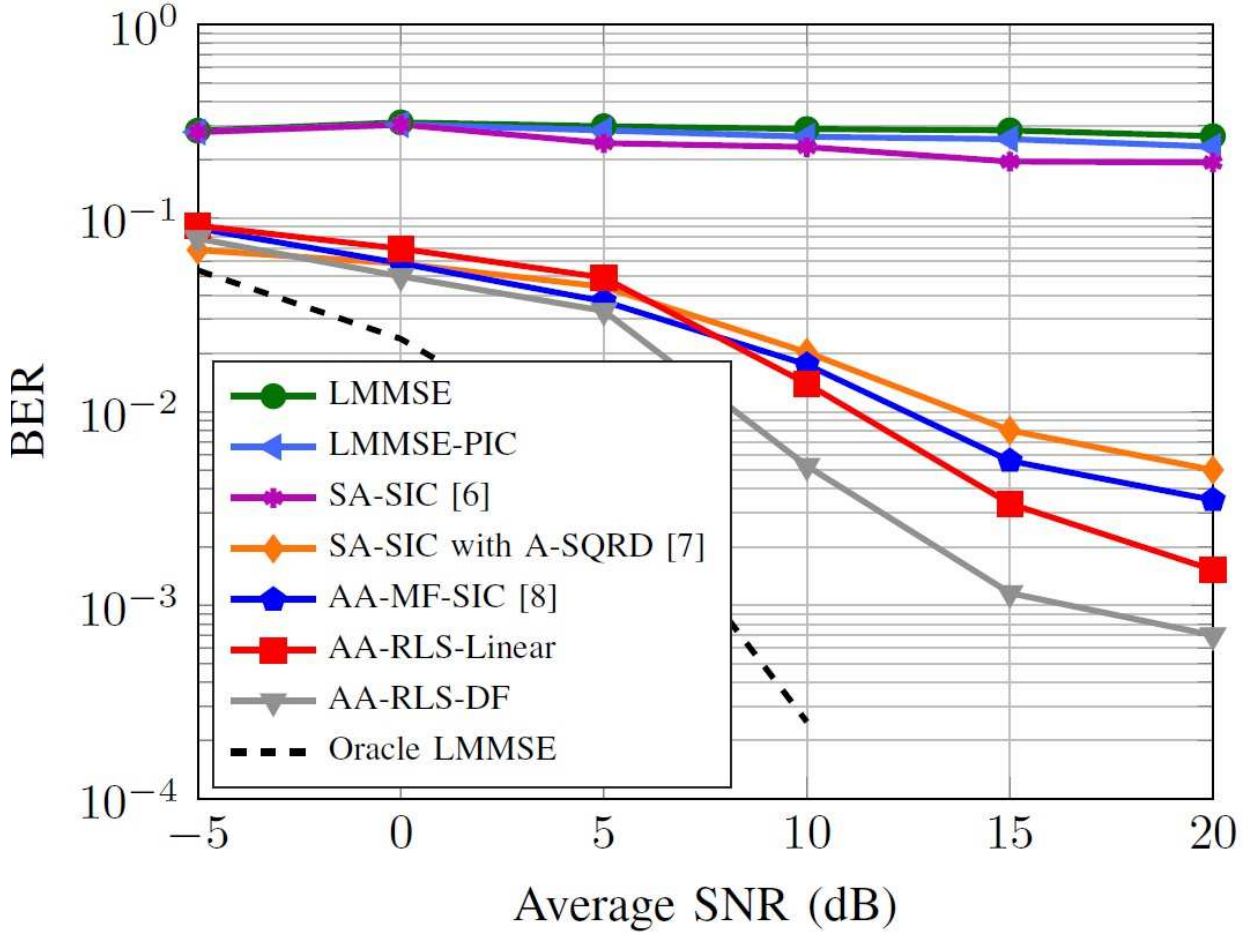}
    \caption{\footnotesize Net symbol error rate vs. Average SNR. Parameters of proposals are $\lambda=0.92$, $\gamma=0.001$ and $\beta=10$, other approaches consider imperfect CSI.}
    \label{fig:result_wout_code}
\end{figure}

\begin{figure}[t]
    \begin{center}
        \includegraphics[scale=0.87]{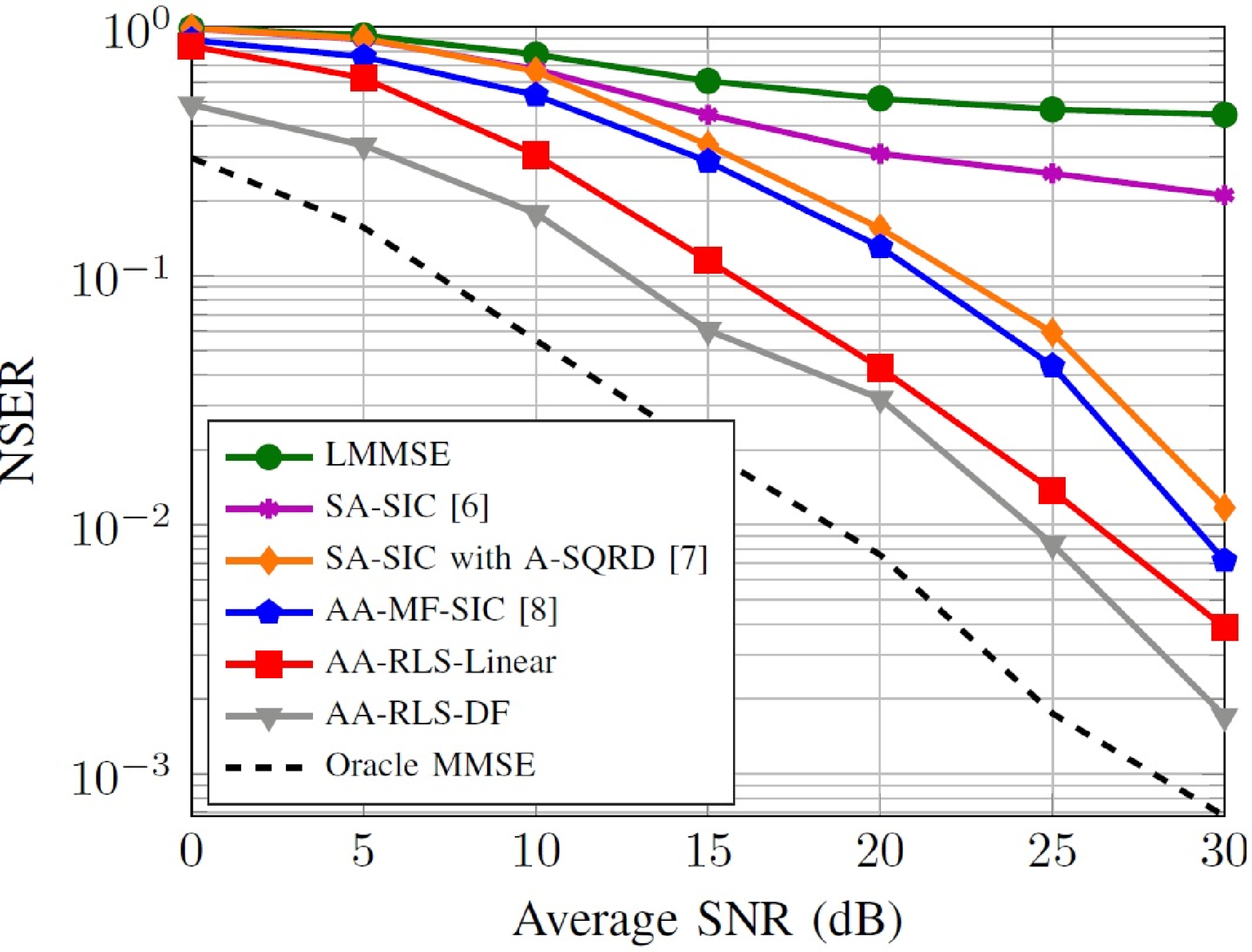}
        \caption{\footnotesize Bit error rate vs. Average SNR. LDPC with block length of 128, symbol rate $R=0.5$,  {refined} by 2 decoding iterations.}
        \label{fig:ldpc}
    \end{center}
\end{figure}

\section{Conclusion}\label{sec:Conc}
We have proposed an activity-aware adaptive DF detector, named AA-RLS-DF, for mMTC scenarios, which unlike competing techniques does not need explicit CSI to detect the symbols. Moreover, an IDD scheme equipped with the AA-RLS-DF technique that exploits device activity to refine the iterative processing has been presented. Simulations have shown that AA-RLS-DF significantly surpass existing approaches.

\ifCLASSOPTIONcaptionsoff
  \newpage
\fi


\end{document}